\documentclass[aps,prd,twocolumn,superscriptaddress]{revtex4-2}

\usepackage[english]{babel}
\usepackage{amsmath, amsfonts, amsthm, amsbsy, amsopn, amsxtra, amscd, amstext}
\usepackage{longtable}
\usepackage{bbm}
\usepackage{enumitem}
\usepackage{graphicx}
\usepackage{xcolor}
\usepackage{wrapfig}
\usepackage{soul}
\usepackage{physics}
\usepackage{amssymb}
\usepackage[hidelinks]{hyperref}

\newcommand\prx{Phys.~Rev.~X}
\def\aj{Astronomical Journal}                 
\def\apj{Astrophysical Journal}                
\def\apjl{Astrophysical Journal}

\def\mnras{MNRAS}            
\def\prd{Phys.~Rev.~D}       
    
\def\cqg{Class.~Quant.~Grav.~}

\newcommand\E[1]{{\left\langle #1 \right \rangle}}
\newcommand\SkipFisher[1]{}

\newcommand\editremark[1]{{\color{red}#1}}
\newcommand\mc{{\cal M}_c}
\newcommand\HideMe[1]{}
\newcommand\unit[1]{{\rm #1}}

\newcommand{\AffiliationCCRG}{
  Center for Computational Relativity and Gravitation, 
  Rochester Institute of Technology, 
  Rochester, New York 14623, USA 
}

\newcommand{\AffiliationUWM}{
  Department of Physics, 
  University of Wisconsin--Milwaukee, 
  Milwaukee, WI 53201, USA 
}

\begin{document}
\title{
   Normal Approximate Likelihoods to Gravitational Wave Events
}
\author{V. Delfavero}
\email[]{msd8070@rit.edu}
\affiliation{\AffiliationCCRG}

\author{R. O'Shaughnessy}
\affiliation{\AffiliationCCRG}

\author{D. Wysocki}
\affiliation{\AffiliationUWM}

\author{A. Yelikar}
\affiliation{\AffiliationCCRG}

\date{\today}

\clearpage{}\begin{abstract}
Gravitational wave observations of quasicircular compact binary mergers in principle provide an arbitrarily complex likelihood
over eight independent ``intrinsic'' parameters: the masses and spins of the two merging objects.
In this work, we demonstrate by example that a multivariate normal approximation over fewer (usually, three) effective
dimensions provides a very accurate representation of the likelihood,
    and allows us to replicate the eight-dimensional posterior over the mass and spin degrees of freedom.
Alongside this paper, we provide the parameters for multivariate normal fits
    for each event published in GWTC-1 and GWTC-2.
The fits are available at \url{https://gitlab.com/xevra/nal-data}.
We demonstrate how these normal approximations provide a highly efficient way to characterize gravitational wave
observations when comparing large numbers of events to detailed formation scenarios.
\end{abstract}
\clearpage{}

\maketitle

\section{Introduction}
The Advanced Laser Interferometer Gravitational Wave Observatory (LIGO) ~\cite{2015CQGra..32g4001T} and  Virgo
\cite{gw-detectors-Virgo-original-preferred,gw-detectors-Virgo-new}
detectors continue  to discover gravitational waves (GW) from coalescing compact binaries consisting of either black holes
(BH) or neutron stars (NS) \cite{LIGO-O1-BBH,LIGO-O2-Catalog,LIGO-O3-O3a-catalog}.
Comparison of these observations to predictions derived from general relativity
produce inferences about each component's mass $m_i$  and dimensionless spin $\pmb{\chi}_i$, collectively denoted by $x$.
From the set of observations and survey selection effects, the underlying population of merging binaries can be
reconstructed hierarchically by, for example, tuning a sufficiently generic phenomenological model to the observations;  see, e.g., 
\cite{gwastro-PopulationReconstruct-Parametric-Wysocki2018,gwastro-PopulationReconstruct-Hierarchical-WysockiDoctor2019,gwastro-PopulationReconstruct-EOSStack-Wysocki2019,LIGO-O2-Rates,LIGO-O3-O3a-RP,2020arXiv200615047T}
and references therein.

Hierarchical inference of the merging binary population relies critically on efficiently calculating \emph{and
  representing} a function that characterizes how well each observation resembles a real quasicircular compact binary GW signal: the 
likelihood ${\cal L}(x)$. 

Customarily,  this function is provided indirectly, via posterior samples $x_k$ that are fair draws from the posterior
  distribution, which is proportional to ${\cal
  L}(x)p(x)$ where $p(x)$ is a fiducial prior distribution, $p(x)$; see, e.g,
\cite{gwastro-PopulationReconstruct-Parametric-Wysocki2018,gwastro-PopulationReconstruct-Hierarchical-WysockiDoctor2019,gwastro-PopulationReconstruct-EOSStack-Wysocki2019}
for a discussion.  

The marginal likelihood $\int {\cal L}(x) p(x|\Lambda) dx$ which measures the agreement between an observation and a proposed
population model $p(x|\Lambda)$ can be evaluated many ways, customarily by weighted Monte Carlo integration using the
provided samples $x_k$. 
In practice, however, this and similar reweighted Monte Carlo integrals can be either inefficient or impractical for many population
    models for two reasons.  On the one hand, the fiducial prior can be poorly suited to the population, producing large
    reweighting factors $p(x)/p(x|\Lambda)$
(e.g., outlier events, narrow population features).
On the other hand, the complementary strategy of deducing a continuous nonparametric approximation to ${\cal L}(x)$ can
likewise introduce computational inefficiency.   For example,  GW190425 has  226,598 associated samples $x_k$ provided
in 
    GWTC-2;  modeling ${\cal L}(x)$ with a kernel density estimate (KDE) based on the posterior samples
    requires a non-trivial amount of time to evaluate that function
    on even a single point.
Similarly, while the likelihood can be evaluated  directly on an unstructured grid and 
    then interpolated to arbitrary $x$ \cite{gwastro-PopulationReconstruct-EOSStack-Wysocki2019,2020MNRAS.499.5972H}, 
this approach generally requires costly and carefully-tuned interpolation.
For these and other reasons, the publicly available samples can be poorly suited
    for downstream  application.

In this work, we demonstrate that a simple normal approximation can very accurately characterize the marginal likelihood
${\cal L}(x)$ for many reported gravitational wave observations.    This approximation is theoretically well-motivated,
being deeply connected to the well-studied Fisher matrix approximation; see, e.g., 
\cite{1995PhRvD..52..848P,2012LRR....15....4J,2008PhRvD..77d2001V,gwastro-mergers-HeeSuk-CompareToPE-Aligned,gwastro-mergers-HeeSuk-FisherMatrixWithAmplitudeCorrections,gwastro-mergers-HeeSuk-CompareToPE-Precessing,2020arXiv201015202B}
and references therein.

This paper is organized as follows.
In Section \ref{sec:methods} we describe our Gaussian approximation to gravitational wave likelihoods,  
\SkipFisher{relate it to
the Fisher matrix,} and describe a procedure to efficiently identify this approximation from posterior samples $x_k$.
We discuss suitable coordinate systems in which to construct a normal approximation.
In Section \ref{sec:results} we employ our  procedure to observations reported in GWTC-1 and GWTC-2  \cite{LIGO-O2-Catalog}\cite{LIGO-O3-O3a-catalog}.
We find a three-dimensional
normal approximation in $\mc,\eta,\chi_{\rm eff}$ very accurately captures most information about each event, including
observations where tides or transverse spins have been modestly constrained.
In Section \ref{sec:applications} we demonstrate the utility of our analysis by directly comparing our GW observations
to the detailed results of a binary evolution simulation's merger rate and mass distribution.
We summarize our results in Section \ref{sec:conclude}.
 \section{Methods}
\label{sec:methods}
\subsection{Gravitational Wave Parameter Inference}
The quasicircular-orbit coalescing compact binaries discussed in this work are fully characterized by their intrinsic
parameters $x$. For black holes, this includes their masses $m_i$ and dimensionless spins $\pmb{\chi}_i$.  
In this paper, we don't consider additional parameters for neutron stars.
To predict the response of a specific detector to the emitted gravitational wave, we also specify seven additional
(extrinsic) parameters specifying the spacetime 4-coordinates of the merger event (e.g., time, distance, and two
coordinates on the sky) and its orientation relative to the Earth (e.g., three Euler angles). 
In this work, we will also use the total mass $M=m_1+m_2$ and mass ratio $q$ defined as $q=m_2/m_1$, assuming $m_1\ge m_2$.
We will also employ two other mass parameterizations: the
 the symmetric mass ratio
 $\eta = (m_1 m_2) M^{-2}$
and the chirp mass  $\mathcal{M}_c = \eta^{3/5} M^{2/5}$.
Finally, we also use an effective spin aligned with the orbital angular momentum, which is consistent with previous work \cite{Damour2001,Racine2009,Ajith2011}
\begin{align}
\chi_{\textrm{eff}} = (\pmb{\chi}_1{m_1} + \pmb{\chi}_2{m_2}) \cdot \hat{\mathbf{L}}/{M}
\end{align}
For clarity, we select a Cartesian coordinate system where $\hat{z}$ is orthogonal
    to the plane of the orbit, such that $\hat{z} = \hat{L}$.

For each set of gravitational wave observations $d$, one can evaluate a full 15-dimensional  likelihood
$\ell(d|x,\theta)$ based on an assumption of Gaussian noise, a measured noise power spectrum, and a forward model for
emitted gravitational waves and the detector response in terms of the intrinsic ($x$) and extrinsic ($\theta$) parameters.  The marginal posterior
distribution $p(x) = [\int d\theta \ell(d|x,\theta)p(x,\theta)]/\int dx d\theta \ell(d|x,\theta)p(x,\theta)$ 
    follows from Bayes' theorem.
In this work, we will assume a fiducial analysis has already been performed
with a fiducial prior $p(x,\theta)$ on these variables, producing a sequence of independent, identically distributed
samples $x_{s},\theta_s$ ($s=1\ldots S$) drawn from a distribution proportional to $\ell(d|x,\theta)p(x,\theta)$.
In this work, we use the preferred posterior samples reported in the published LIGO catalogs for GWTC-1 \cite{LIGO-O2-Catalog}
and GWTC-2 \cite{LIGO-O3-O3a-catalog}.  
These calculations employ a standard prior, equivalent to the expressions provided below  \cite{lange2018rapid}.
The mass prior is uniform in component \emph{detector-frame} masses, equivalent to the following expression versus
detector-frame chirp mass $\mathcal{M}_{c,z}=(1+z)\mathcal{M}_c$ and $\eta$:
\begin{equation}\label{eq:simple_prior}
p(\mathcal{M}_{c,z}, \eta)d{\mathcal M}_{c,z}d\eta = \frac{4}{(M_{\mathrm{max}} - m_{\mathrm{min}})^2} \frac{\mathcal{M}_{c,z}d{\mathcal M}_{c,z}d\eta}{\eta^{6/5}\sqrt{1 - 4\eta}}
\end{equation}
The spin prior is uniform in spin magnitude, $\chi_i$, and isotropic in orientation.

Our method can be applied to samples representing any waveform model, and will have results consistent with such a waveform.
These posterior samples are independent, identically distributed random draws
from a distribution $\propto \ell (x,\theta)p_{\rm ref}(x,\theta)$ where $p(x,\theta)$ is a  fiducial prior distribution
over extrinsic variables $\theta$ and intrinsic source-frame variables $x$.   From these samples, we seek to estimate
the marginal likelihood ${\cal L}(x) = [\int d\theta \, \ell(x,\theta)p_{\rm ref}(x,\theta)]/p_{\rm ref}(x)$ versus the
source-frame intrinsic parameters $x$, where $p_{\rm ref}(x)$ is a fiducial prior for the source frame parameters.  [We
  emphasize that $p_{\rm ref}(x)$ is not precisely equal to the marginal distribution of $p_{\rm ref}(x,\theta)$,
  insofar as detector-frame and source-frame masses differ.]   We
construct our estimate $\hat{\cal L}(x)$ up to an overall normalization constant using a weighted kernel density estimate:
\begin{align}
\hat{q}(x) = {\cal K} \sum_k w_k G(x-\mu,\Sigma_S)
\end{align}
where $G$ is a multivariate normal distribution with covariance $\Sigma_S$
    expressed in $\mc,\eta,\chi_{\rm eff}$ coordinates,
    and where $\Sigma_S$  and ${\cal K}$ are
    empirically determined constants that ensure
    $\hat{q}(x)$ approximates ${\cal L}(x)$,
    and that these quantities are normalized consistently
    over the domain of $x$, including edges.
Specifically, we adopt
\begin{align}
w_k= \frac{1}{p_{\rm ref}(x)} = \frac{1}{p(\mathcal{M}_{c,z}, \eta) (1+z) p(\chi_{\textrm{eff}}|\eta)}\nonumber \\
=\frac{1}{p(\mathcal{M}_{c}, \eta) (1+z)^2 p(\chi_{\textrm{eff}}|\eta)},
\end{align}
where $p(\mathcal{M}_{c,z}, \eta)$ is given  by Eq. (\ref{eq:simple_prior}). 
The prior $p(\chi_{\rm eff}|\eta)$ is the  inferred marginal distribution of $\chi_{\rm eff}$ at fixed $q,\mc$, derived
assuming a uniform-magnitude/isotropic spin prior; see Appendix B of \cite{2021arXiv210409508C}
 for a concrete expression.
In effect, this choice re-weights the posterior samples to a uniform prior in $\mc,\eta$ and $\chi_{\rm eff}$, so the
weighted KDE is the (marginal) likelihood in this space.  
In this work, we retain the fiducial distance prior ($p(d_L)\propto d_L^2$); however, we emphasize that the choice of
distance prior propagates directly into any inferred Gaussian which does not include distance among its parameters, and that these calculations must be repeated with an
appropriately modified weight if a different distance prior is desired.\footnote{We also find that a multivariate gaussian in
  detector-frame parameters augmented by $1/d_L$ provides an excellent approximation to the corresponding posterior, and
  is useful in applications where a
  range of source redshift distributions arise.} 
We choose a covariance matrix $\Sigma_S$ based on the covariance of the data, using conventional bandwidth-selection rules for
kernel density estimates (Scott's rule); see, e.g.,
\cite{Merritt1994,ThomsonTapia,mm-statistics-nonparametric-timeseries-RamsaySilverman} and references therein.

Events near equal mass have the unfortunate condition that the KDE will fall
    off rapidly near $\eta = 1/4$, which is a hard limit on the 
    symmetric mass ratio, $\eta$.
To account for this, we mirror the posterior samples about $\eta = 1/4$
    when generating the KDE for such events.
We never sample from the region where $\eta > 1/4$, so this method
    is justified.

\subsection{Multivariate Normal Approximations}
A  multivariate normal approximation  is a customary and effective approximation for the gravitational wave likelihood,
in suitable coordinates  and accounting for the finite domain of $x$; see, e.g., \cite{NRPaper,1995PhRvD..52..848P,gwastro-mergers-HeeSuk-CompareToPE-Aligned,gwastro-mergers-HeeSuk-FisherMatrixWithAmplitudeCorrections,gwastro-mergers-HeeSuk-CompareToPE-Precessing}.  
Conversely, a multivariate normal approximation does not by itself arise in generic coordinates, but only in parameters
particularly well-adapted to the signal, notably including (at low mass) the binary chirp mass $\mc$, symmetric mass ratio $\eta$, and
effective spin $\chi_{\rm eff}$.  In other coordinate systems, a likelihood well-approximated by a multivariate normal will appear
non-Gaussian.  Similarly, a multivariate normal approximation by itself does not account for the finite domain of $x$:
most notably, $\eta\le 1/4$, so the natural multivariate normal for comparable-mass binaries is invariably truncated by
the equal-mass line.

In this section, we describe how to recover the parameters 
    $\mathbf{\mu},\mathbf{\Sigma}$ 
    of a multivariate normal approximation
$G(x-\mathbf{\mu},\mathbf{\Sigma}) = ( |2\pi\mathbf{\Sigma}|)^{-1/2} \exp [-1/2(x-\mathbf{\mu})^T \mathbf{\Sigma}^{-1}(x-\mathbf{\mu})]$.

We parameterize our multivariate Gaussian with vectors components
    for the mean $\mathbf{\mu}$ and standard deviations $\mathbf{\sigma}$,
    and the symmetric correlation matrix $\mathbf{\rho}$.
The composition of the covariance from our parameterization is therefore 
    $\mathbf{\Sigma} = \mathbf{\sigma} \mathbf{\rho} \mathbf{\sigma}$.
Alternatively, 

\begin{equation}\label{eq:cov_expansion}
\mathbf{\Sigma} = 
\begin{pmatrix}
    \sigma_1^2 && \sigma_1\sigma_2 \rho_{1,2} && \cdots && \sigma_1\sigma_N\rho_{1,N} \\
    \sigma_1\sigma_2\rho_{1,2} && \sigma_2^2  && \cdots && \sigma_2\sigma_N\rho_{2,N} \\
    \vdots && \vdots && \ddots && \vdots \\
    \sigma_1\sigma_N\rho_{1,N} && \sigma_2\sigma_N\rho_{2,N}  && \cdots && \sigma_N^2 \\
\end{pmatrix}
\end{equation}

For a $N$-dimensional variable space, there are $N$ mean parameters, $\mathbf{\mu}$,
    $N$ variance parameters, $\sigma(\mu)$,
    and $(N^2 - N)/2$ correlation parameters, $\mathbf{\rho}$.
Therefore, there are a total of $(N^2 + 3N)/2$ parameters.

For many problems, the mean and covariance of a set of data will
    provide parameters for a well-fitting Gaussian.
We see that this is not true for nearly any events in the coordinates
    used for parameter estimation.
Specifically, the symmetric mass ratio, $\eta$, approaches a limit
    of $1/4$ at equal mass.
In this coordinate, the mean and covariance of the data will almost
    never provide parameters for a well-fitting Gaussian,
    because the limit imposed by the coordinates restricts the range
    of the data.
We see that using the mean and covariance to construct parameters for our
    Gaussians will introduce a bias away from equal mass, for all but
    the most unequal mass events.

A Gaussian can still fit the likelihood, but the parameters 
    can no longer be calculated directly from the mean and covariance
    of the data.
We optimize the parameters of each Gaussian using
    simulated annealing \cite{simulated-annealing}, because it scales well
    as the dimensionality of our data increases.

In order to select these parameters in a more reliable way, 
    we defined an objective function which minimized the difference between this Gaussian
    approximation $G(x-\mu,\Sigma)$ and a
    Gaussian KDE.
Specifically, if $\hat{q}$ is our estimate for the
    KDE and $G$ is our estimate for the Gaussian, we try to minimize $\sum_k |\hat{q}(x_k)-G(x_k-\mu,\Sigma)|^2$. 
In this expression, the points $x_k$ are chosen by randomly drawing from the distribution defined by $\hat{q}$, 
    with the added requirement that $x_k$ lie within the finite range.
This added requirement reduces the error caused by the finite range.
We use simulated annealing to optimize this objective function over the parameters describing our mean $\mu$ and
covariance $\Sigma$. 

\subsection{Measuring Similarity}

To assess the reliability of our Gaussian approximations, 
    we use the Kullback--Leibler divergence $D_{\mathrm{KL}}(p\parallel q) = \int dx \, p(x) \ln
p(x)/q(x)$ applied to various one, two, and three-dimensional marginal distributions
\cite{mm-stats-Dkl-def}.  
For example, we compute the KL
    divergence between each one-dimensional marginal distribution $D_{\mathrm{KL}}(p_k\parallel q_k)$ for the parameters
$k=\{\mc,\eta,\chi_{\rm eff}\}$, and in particular the root mean square of all these KL divergences:
\begin{align}
\bar{D}_{\mathrm{KL}}  = \sqrt{\sum_{k} D_{\mathrm{KL}}(p_k \parallel q_k)^2}
\end{align}
In these expressions, the  $p_k$ represent our (truncated) one-dimensional Gaussian distributions for each parameter in
\emph{likelihood},
while $q_k$ represent our KDE-based estimate for the likelihood.
This is done in each dimension instead of using the joint three-dimensional distribution
    because of binning difficulties with a three-dimensional KL divergence.

To demonstrate the sufficiency of our three-dimensional Gaussian approximation to enable complete recovery of the posterior,
    we use RIFT \cite{lange2018rapid} to generate a synthetic (intrinsic, source-frame) posterior under the assumption
that the marginal likelihood is characterized by the Gaussian approximation described above.
We then construct Jensen-Shannon (JS) divergences
    between the true posterior and our approximation,
    for intrinsic parameters reflecting mass as well as 
    aligned spin degrees of freedom \cite{MENENDEZ1997307}.

\section{Results}
\label{sec:results}
\begin{figure}
\includegraphics[width=\columnwidth]{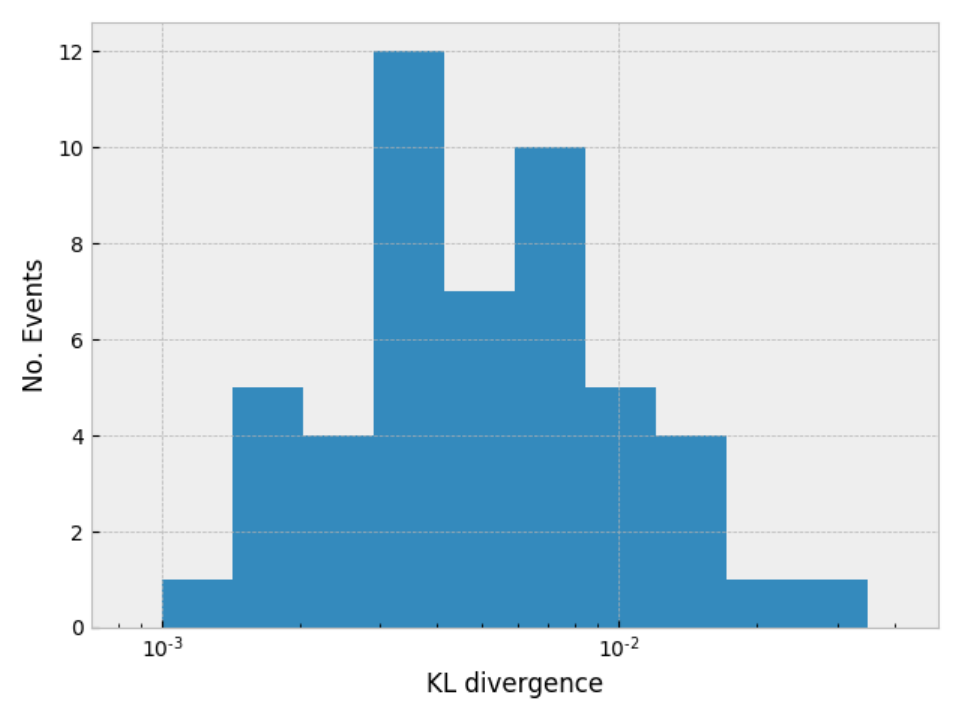}
\caption{\label{fig:Diagnostics} Histogram of KL divergences between our three-dimensional KDE $\hat{L}(x)$ and the inferred Gaussian approximation.
    GWTC-1 events are modeled using SEOBNRv3 (except GW180817, which uses IMRPhenomPv2NRT\_lowSpin).
    GWTC-2 events are modeled using the preferred catalog samples.}
\end{figure}

\begin{figure*}
\includegraphics[width=0.32\textwidth]{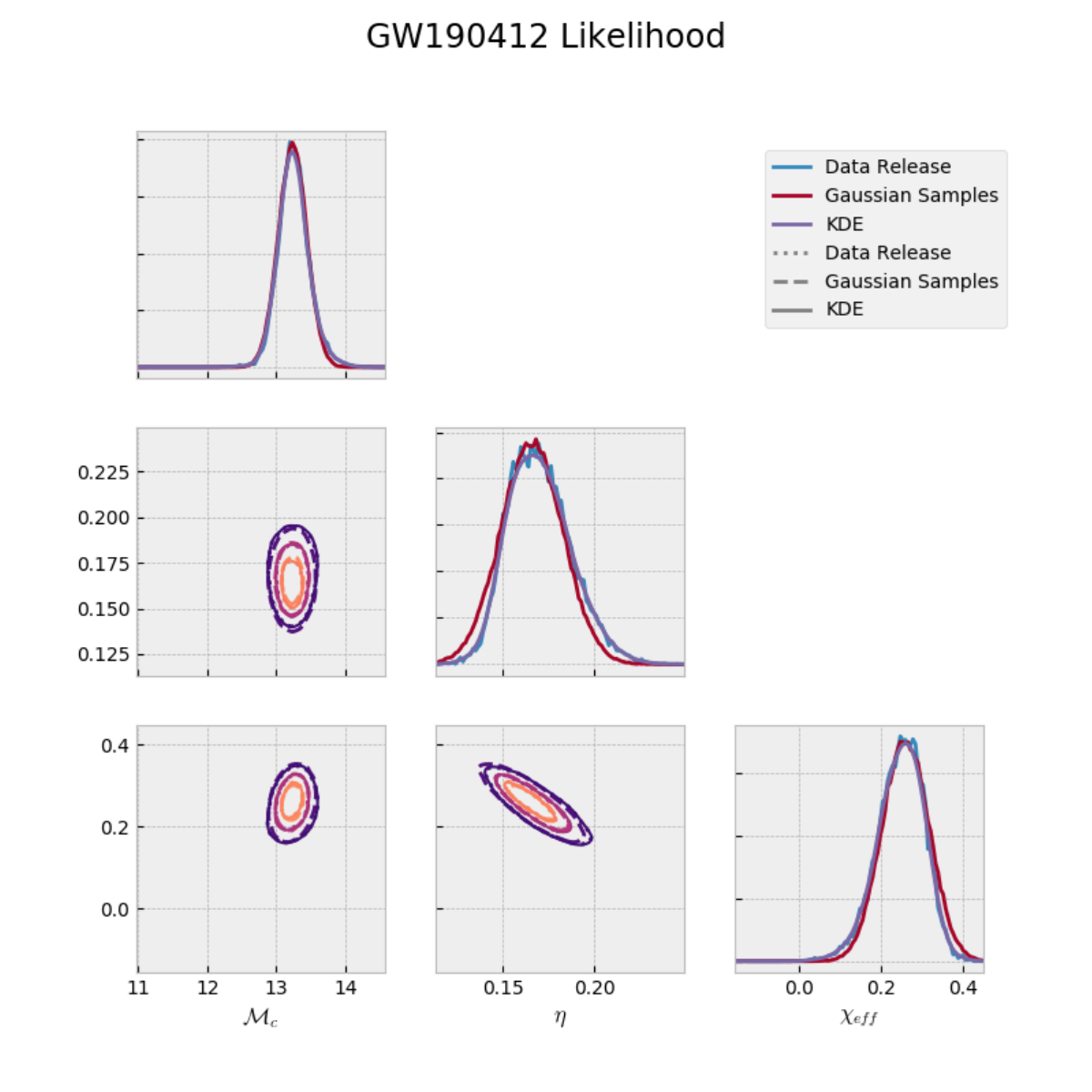}
\includegraphics[width=0.32\textwidth]{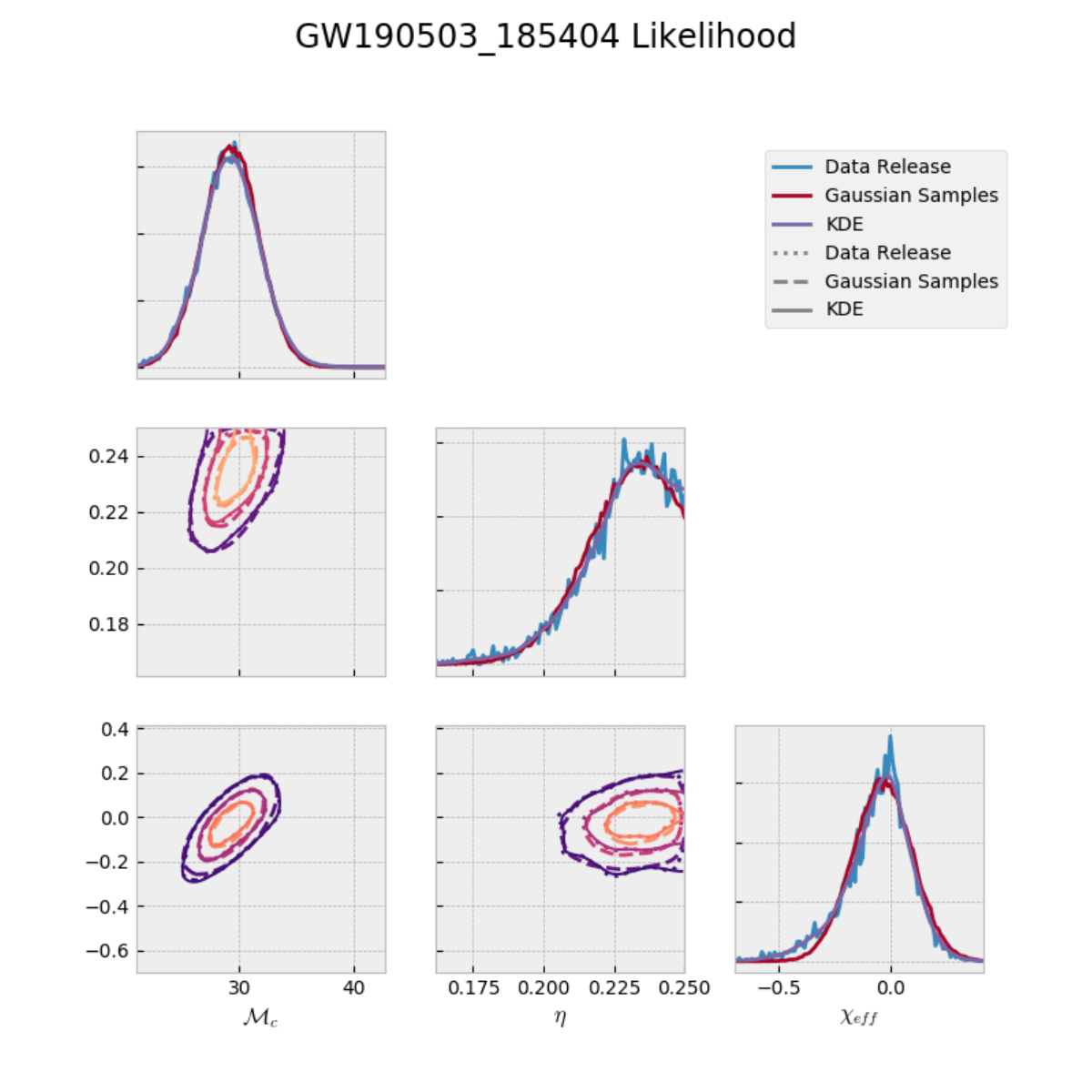}
\includegraphics[width=0.32\textwidth]{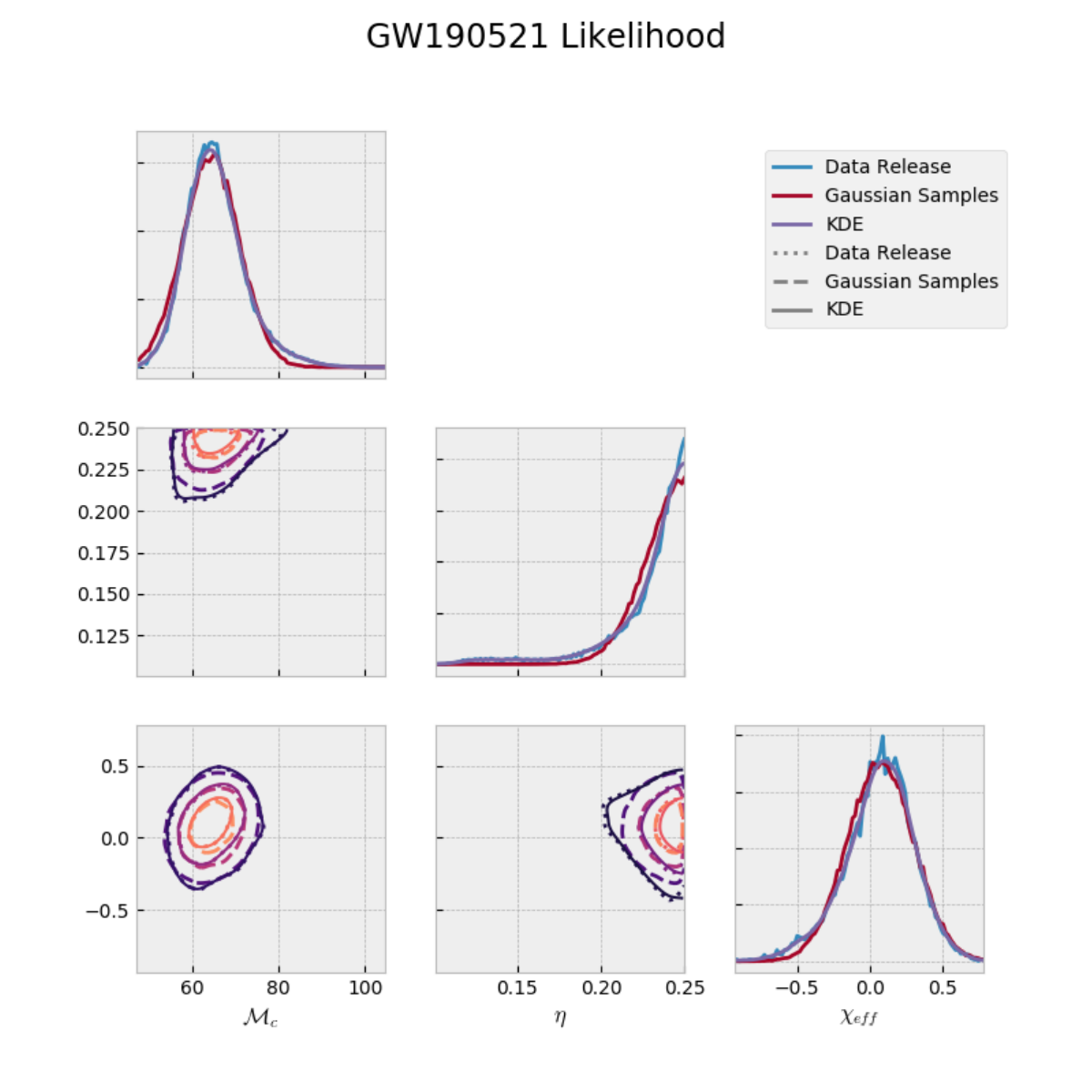}

\includegraphics[width=0.32\textwidth]{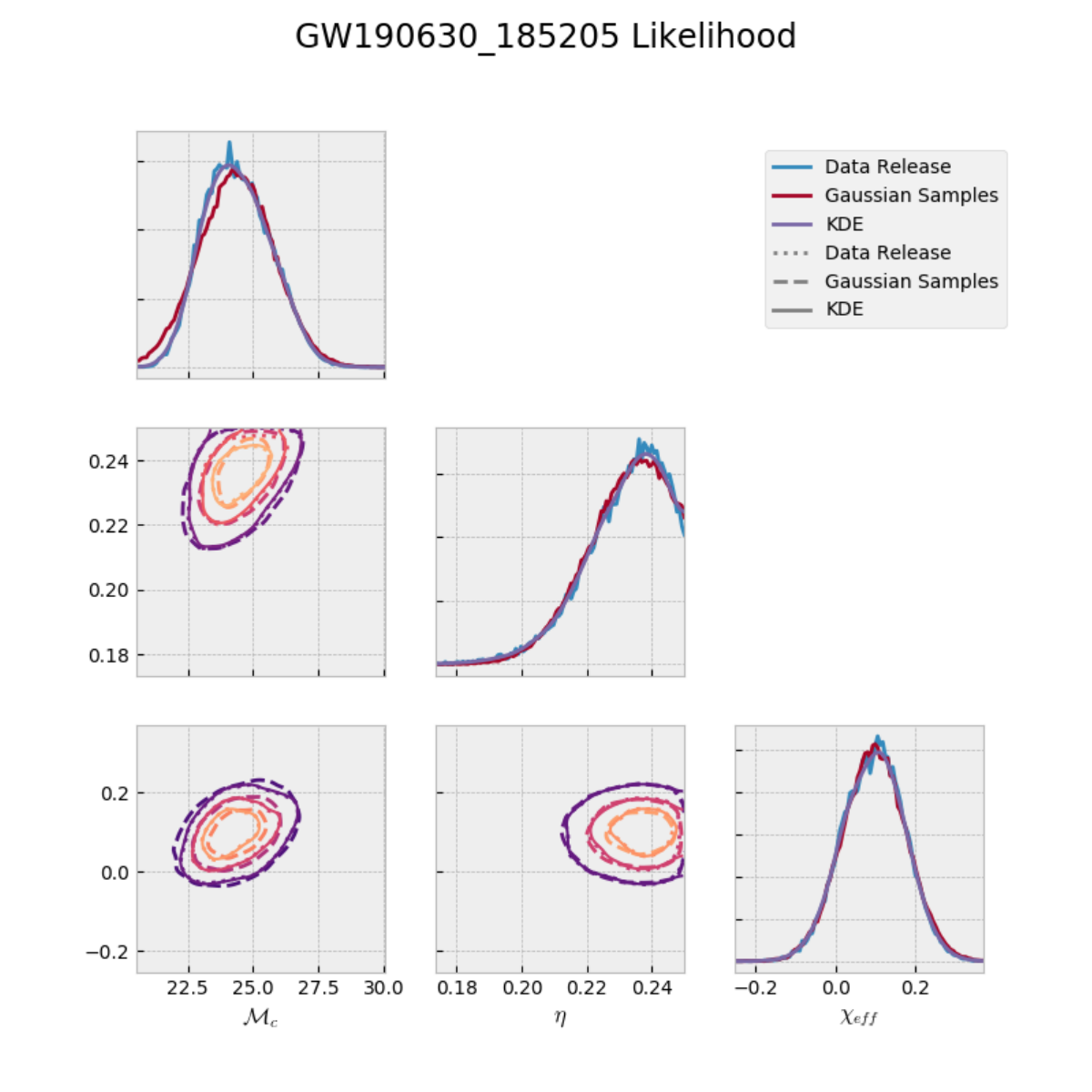}
\includegraphics[width=0.32\textwidth]{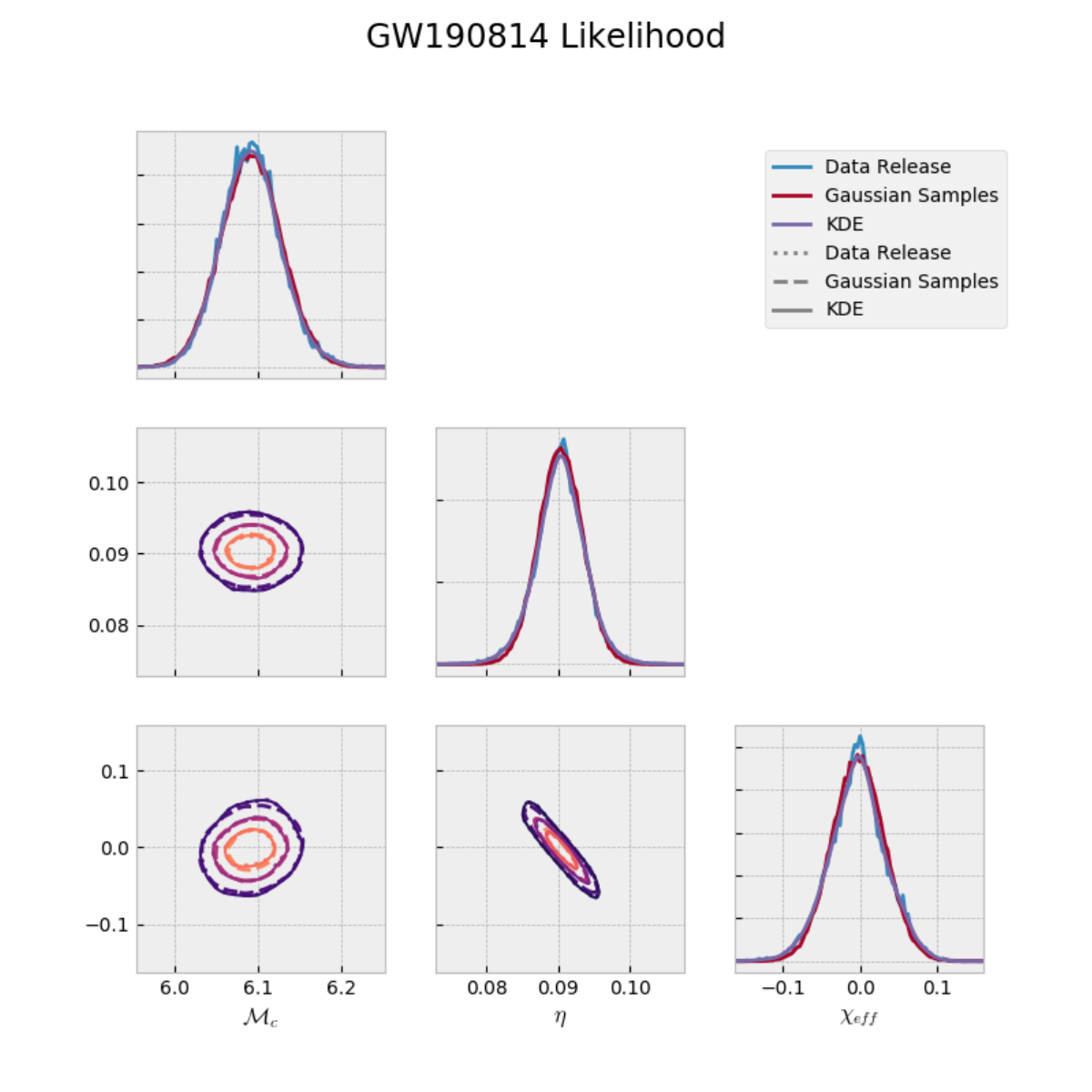}
\includegraphics[width=0.32\textwidth]{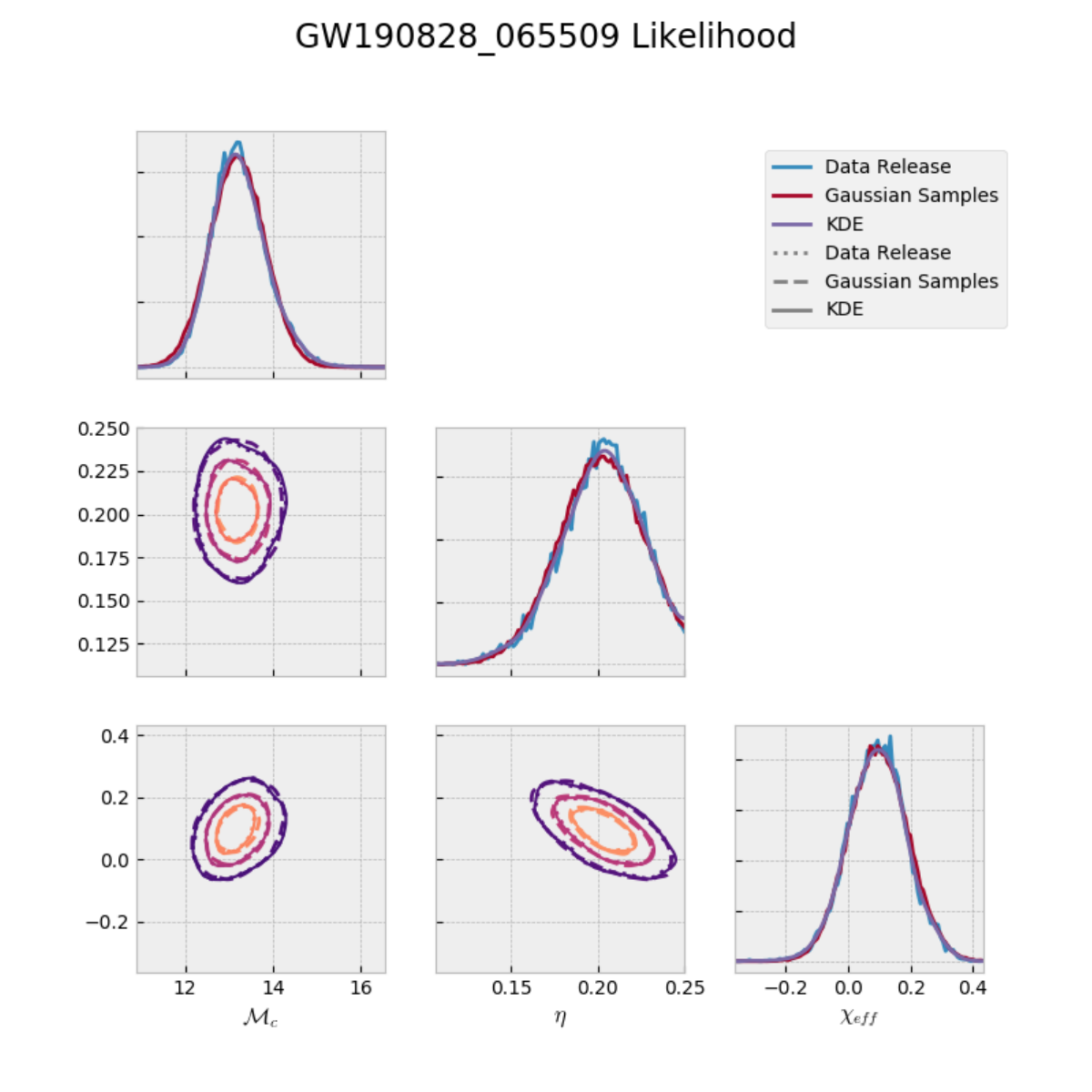}
\caption{\label{fig:ExamplesPrimary}
    \textbf{Illustrations of Gaussian fits for interesting events} (e.g., where the peak likelihood is offset from equal mass).
    The one-dimensional  figures along the diagonal of each corner plot show the inferred marginal likelihood Gaussian approximation (red), KDE approximation (purple), and
    reweighted original posterior samples (blue), where the reweighting converts to a uniform prior in
    $\mc,\eta, \chi_{\rm eff}$. 
    The two dimensional marginalizations show the 25\%, 50\%, and 75\% contour
        for the reweighted catalog (dotted), the KDE (solid), and the Gaussian (dashed).
 These events  are all well-fit,
        including the high-mass event GW190521; see Figure \ref{fig:Diagnostics} for quantitative measures of similarity.
}
\end{figure*}
\begin{figure*}
\includegraphics[width=0.32\textwidth]{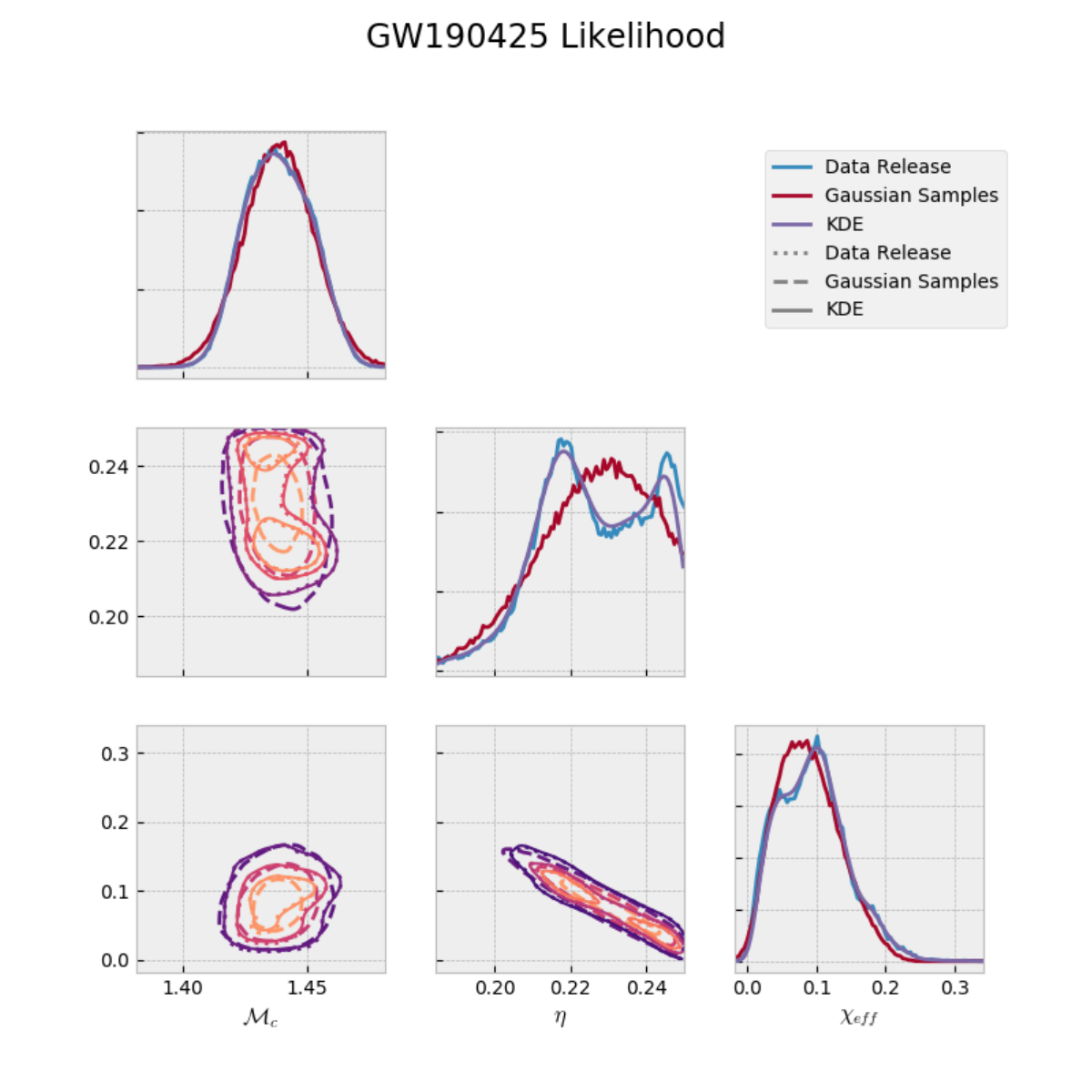}
\includegraphics[width=0.32\textwidth]{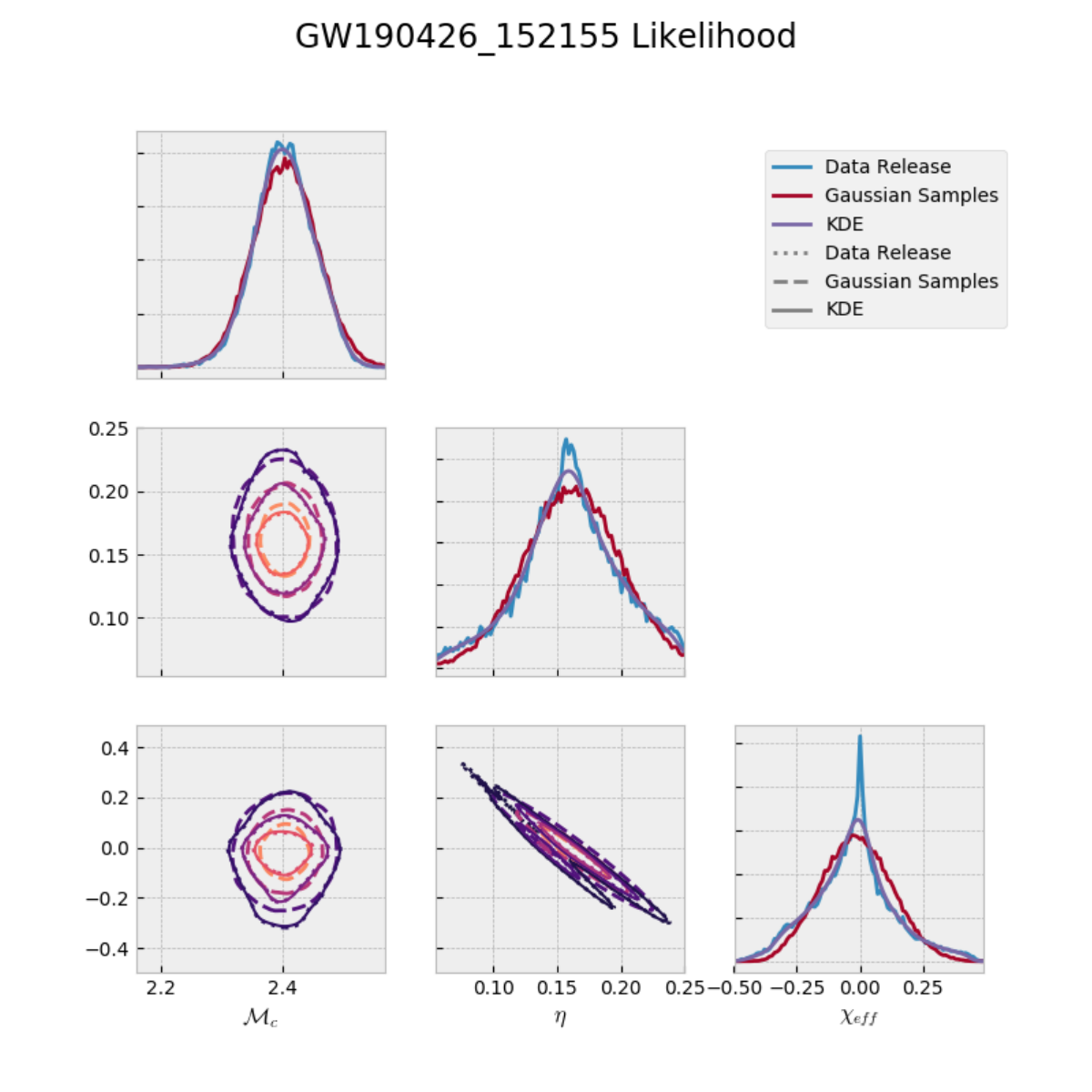}
\includegraphics[width=0.32\textwidth]{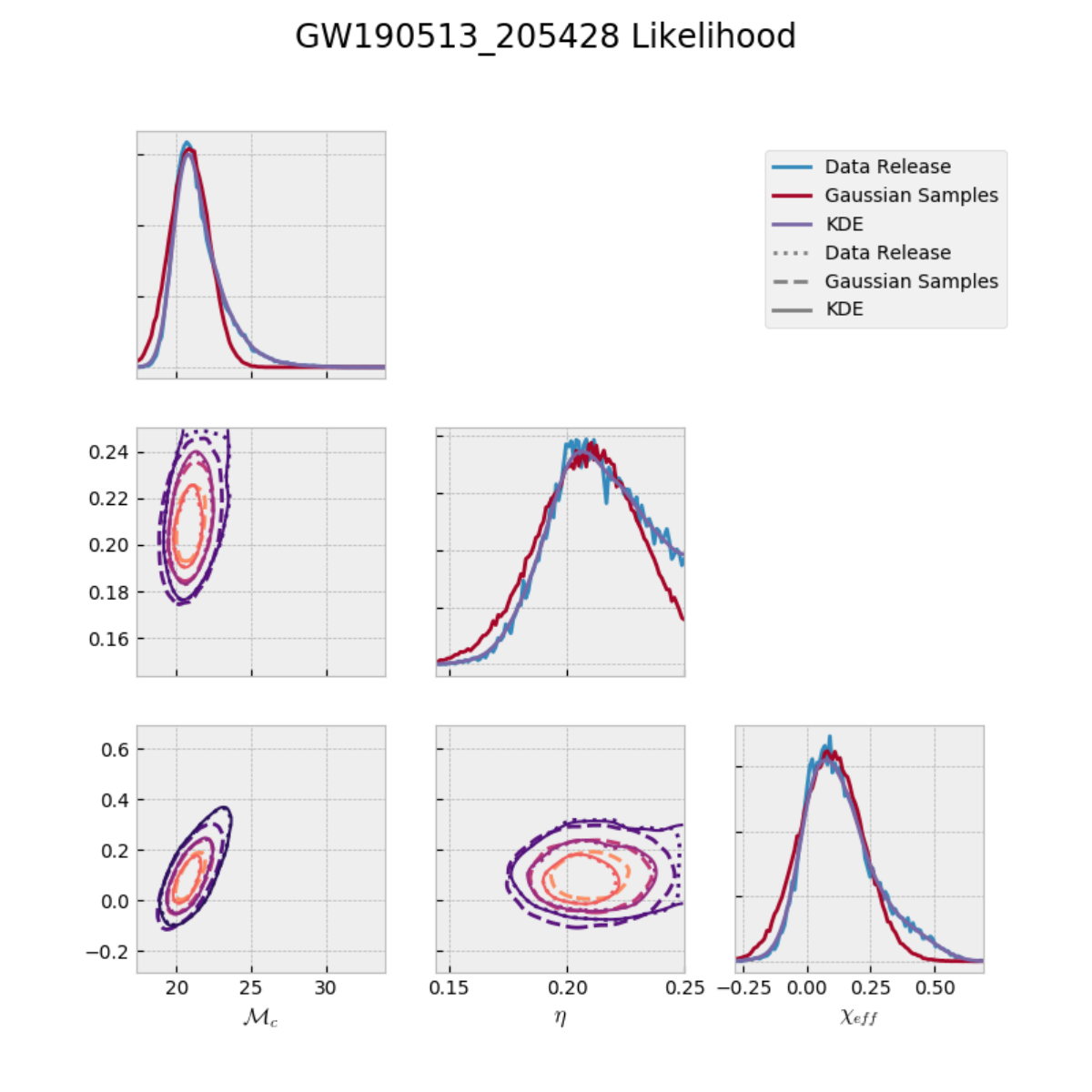}

\includegraphics[width=0.32\textwidth]{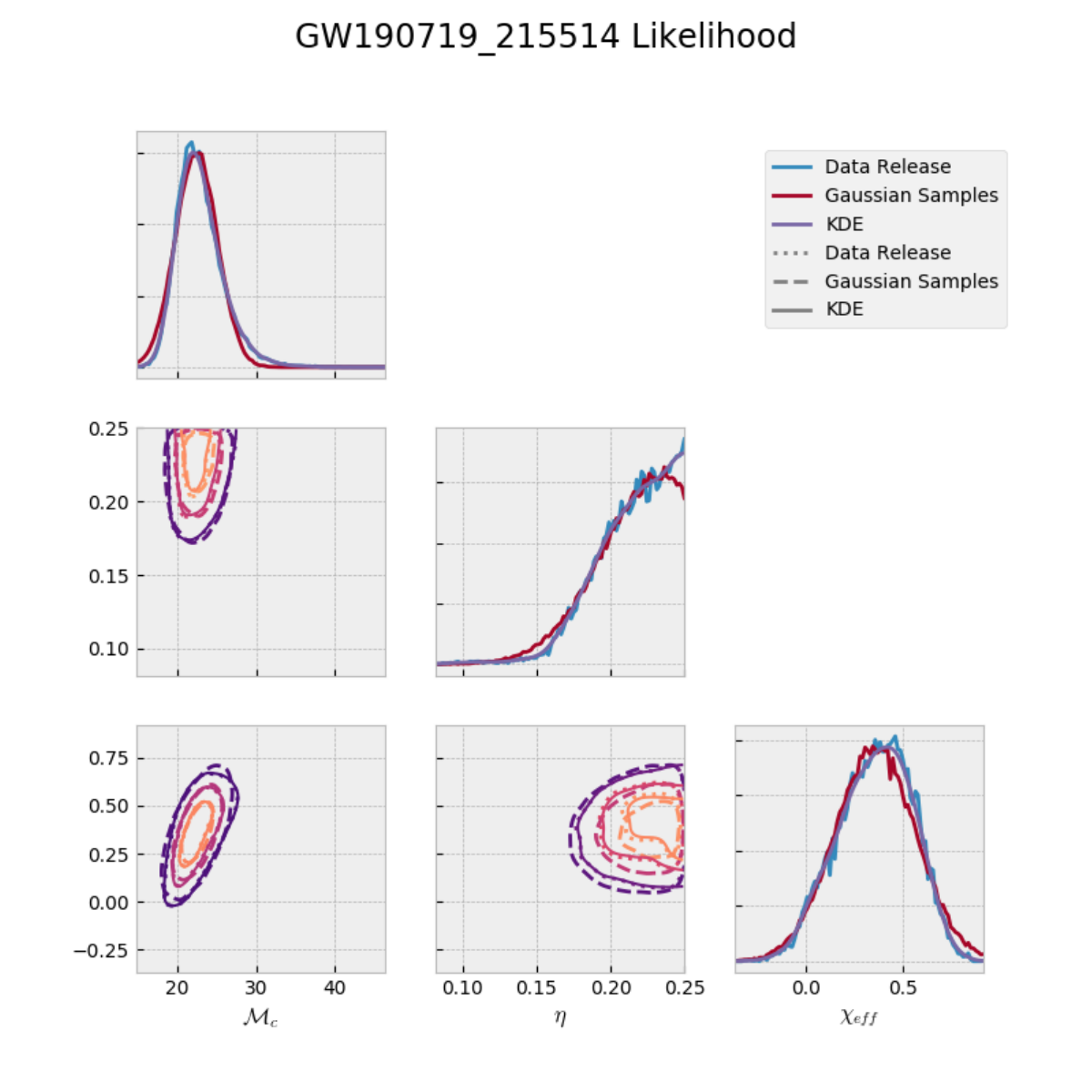}
\includegraphics[width=0.32\textwidth]{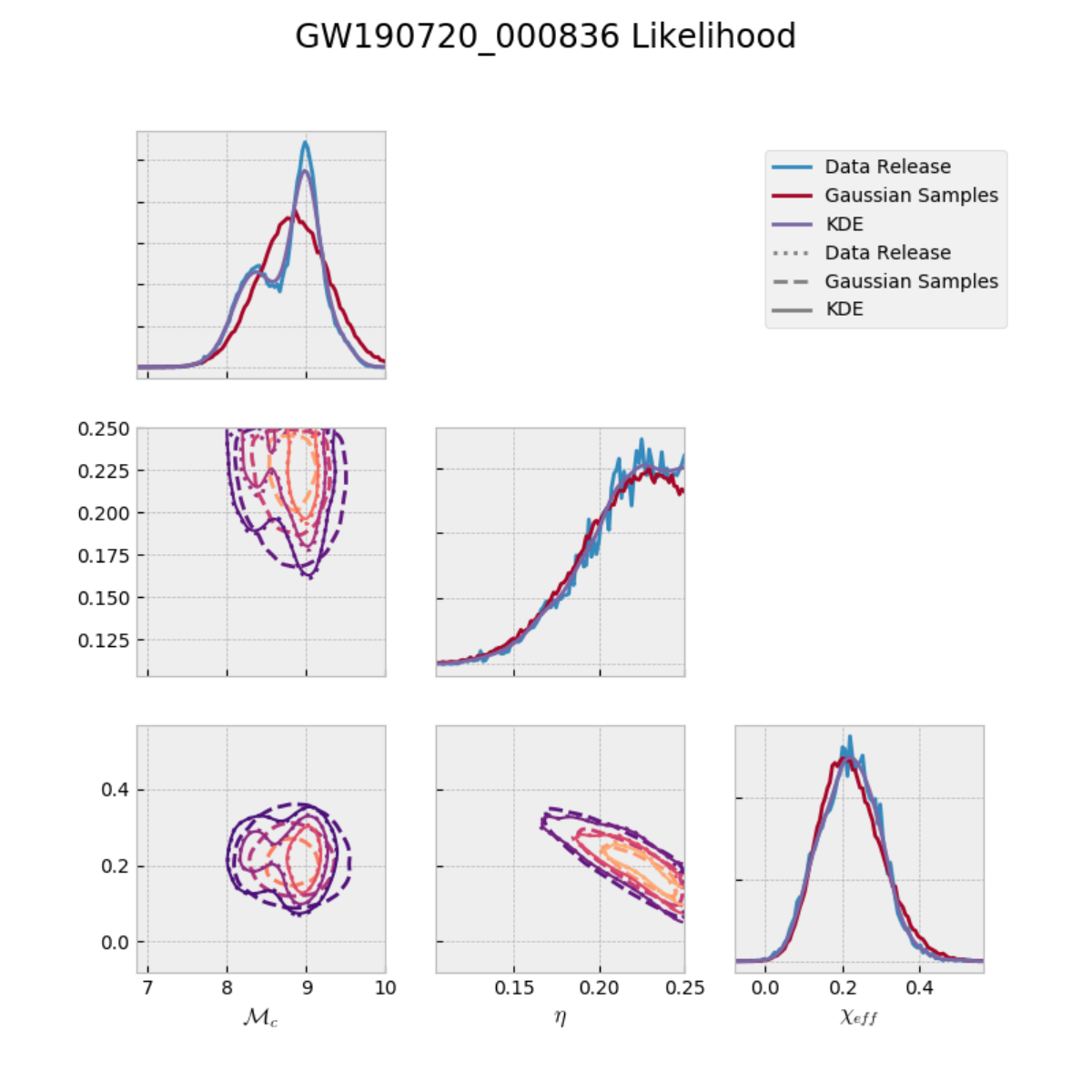}
\includegraphics[width=0.32\textwidth]{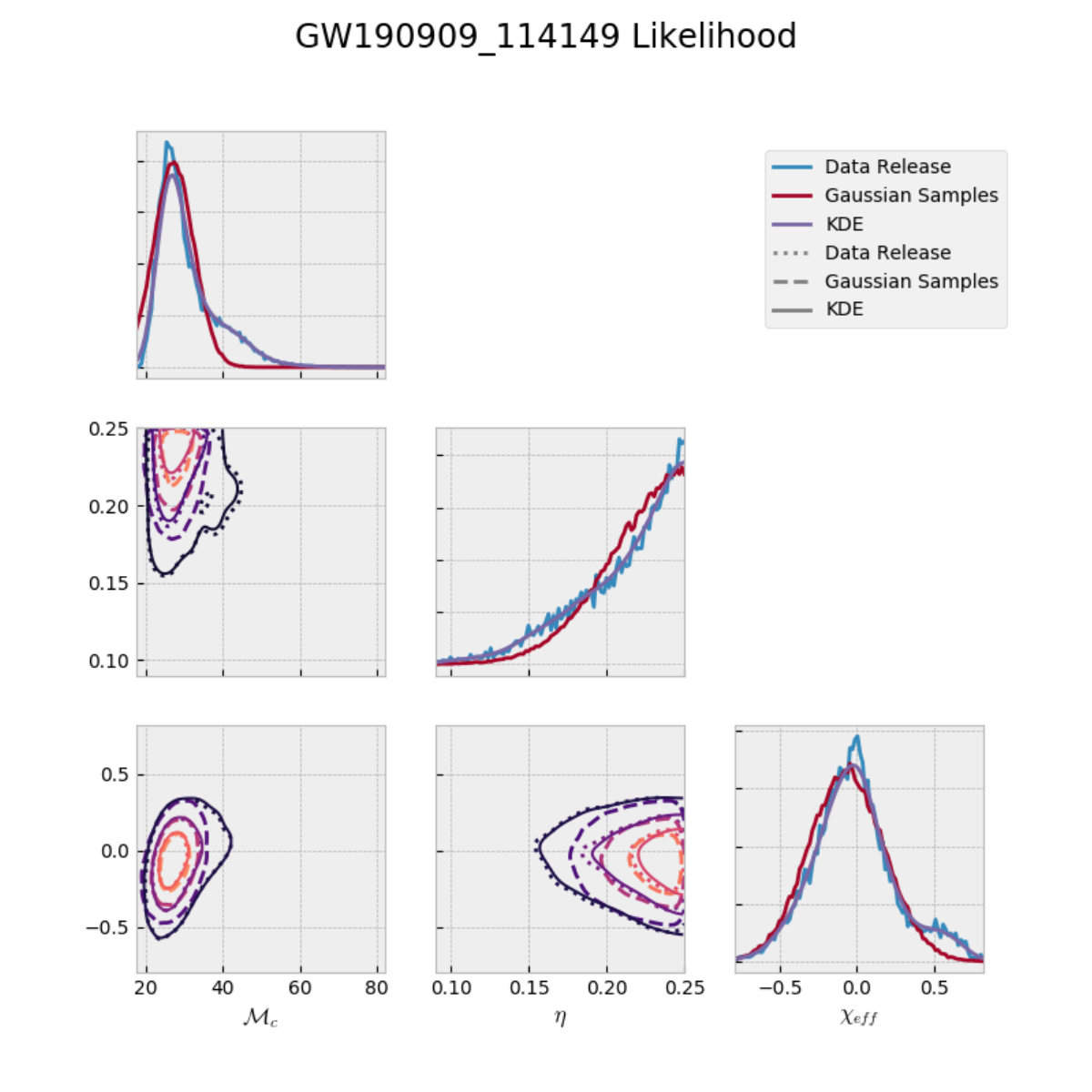}
\caption{\label{fig:ExamplesSecondary}
    \textbf{Selected Gaussian fits which omit some  features of
        the estimated likelihood.}
     The two cases of multimodal  likelihoods are not well-fit by a single Gaussian, as are cases with  a sharp spike or extra lump.
    GW190719 is interesting, in that one might expect this binary to have
        equal mass by visual inspection,
        but the peak observed in the fit is offset from $\eta=1/4$.
    Still, the Gaussian models presented
        boast a convincing goodness of fit estimate 
        (Figure \ref{fig:Diagnostics}), and the offsets are small compared to statistical and astrophysical uncertainty.
}
\end{figure*}

\begin{figure}
\includegraphics[width=\columnwidth]{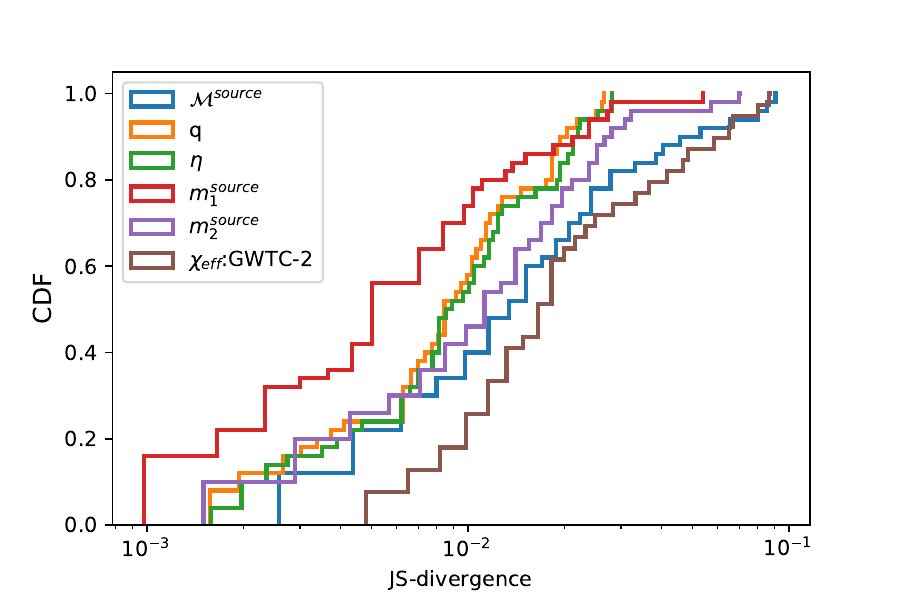}
\caption{\label{fig:Diagnostics:4d}Cumulative distribution of JS divergences between input posterior distribution and
        one regenerated from the (three-dimensional) Gaussian likelihoods provided in this work.}
\end{figure}

We have evaluated our multivariate Gaussian approximations for all events, using 
    GWTC-1 and GWTC-2 \cite{LIGO-O2-Catalog,LIGO-O3-O3a-catalog}.
For GWTC-1, SEOBNRv3 was used for each event apart from GW170817, for which IMRPhenomPv2NRT\_lowSpin was used.
For GWTC-2, the preferred samples from the catalog were used for each event.
The parameters of these Gaussians are available at \url{https://gitlab.com/xevra/nal-data}.
Figure \ref{fig:Diagnostics} provides quantitative diagnostics of our goodness-of-fit for all events reported to date,
for the three-dimensional space of fitting parameters.
For almost all events, our multivariate Gaussian approximations provide an excellent approximation to the
three-dimensional marginal distribution in $\mc,\eta,\chi_{\rm eff}$.   As an illustration, 
Figure \ref{fig:ExamplesPrimary} compares our Gaussian approximations to the inferred three-dimensional marginal likelihood
$\hat{L}$ for six selected events, including representative massive and low-mass compact binaries.
For most events, the peak of our Gaussian approximation is consistent with equal mass; 
    however, for many events, the Gaussian approximation is notably offset from the equal-mass line:  
        GW170104, GW170809, GW190412, GW190426c, GW190503bf, GW190512at, GW190513bm, GW190519bj, 
        GW190521r, GW190630ag, GW190706ai, GW190814, GW190828j, GW190929d.
Given the fifty observations we focus on in this study,
    we would expect to find on average 2.5 events at least two standard
    deviations away from equal mass, 
    even under the assumption that most events
    have mass ratios indistinguishable from unity.
We find nine (three) events with mass ratios offset by at least
    one (two) standard deviation(s), 
    which is consistent with that assumption.

Because our approximation works so well, we highlight in Figure \ref{fig:ExamplesSecondary} the few cases where our Gaussian
    approximation does not capture key features of the reported fiducial posterior.
Only a handful of events at the lowest masses and/or associated with low SNR have features which are manifestly non-Gaussian in these three
variables.    As seen in Figure \ref{fig:ExamplesSecondary}, only two events show notable multimodality in the inferred likelihood: the low-amplitude low-mass event
    (GW190425) and the oddly-behaved fiducial PE for GW190720.
Even in these cases which proved difficult to fit, the Gaussian model produced 
    has a low KL divergence relative to directly inferred Gaussian parameterizations,
    and we provide examples of few key events (Figure \ref{fig:Diagnostics}).

To further demonstrate the utility of these extremely simple approximations, we regenerate full six-dimensional
posterior distributions, assuming the marginal likelihood only depends on $\mc,\eta,\chi_{\rm eff}$ according to our
Gaussian approximation.  We otherwise adopt comparable priors for mass and spin: uniform in (source-frame) component
masses, uniform in component spin magnitudes, and isotropic in component spin directions.  Figure \ref{fig:Diagnostics:4d} shows quantitative comparisons between our input posterior distribution and the posterior
inferred from our Gaussian approximation.
 Precisely insofar as previous gravitational wave observations don't constrain transverse
spin, this three-dimensional likelihood approximation allows us to regenerate a full eight-dimensional posterior distribution which agrees extremely
well with the full posterior distribution used to generate it.  
At very low mass, however, gravitational wave observations can and do modestly  constrain transverse spins, so our
three-dimensional approximation increasingly breaks down.  
In subsequent work, we will report on the agreement
between ${\cal L}$ with more intrinsic and few
extrinsic degrees of freedom (e.g., with all eight spin degrees of freedom, distance, and tides). 
However, for most applications of observations through GWTC-2,
    the simple three-dimensional approximation suffices.

\section{Applications}
\label{sec:applications}
\begin{figure}
\includegraphics[width=\columnwidth]{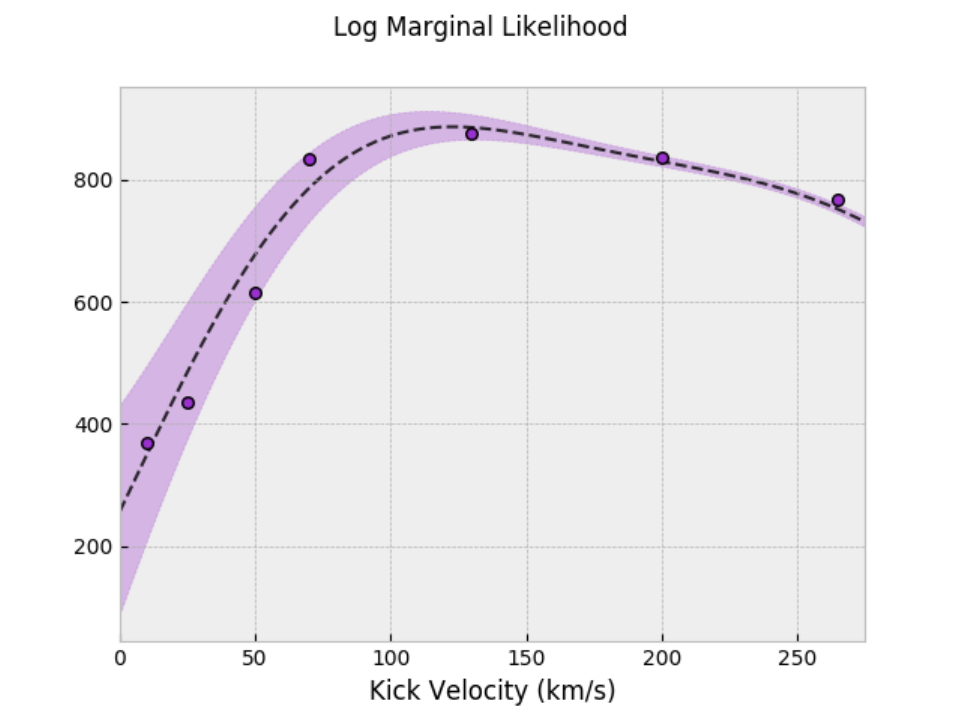}
\caption{\label{fig:Application} 
The likelihood that StarTrack simulations (points)
    represent physical assumptions,
    as a function of supernova natal recoil kick velocity.
The dotted line is a Gaussian Process Regression (GPR)
    interpolation of this likelihood,
    and the region of one sigma is shaded.
}
\end{figure}

One application of our fast normal approximate likelihoods is to quantitatively compare detailed  compact binary
formation models to gravitational wave observations.    
Simple parameterized phenomenological models can be efficiently compared to GW observations using previously-reported
samples  \cite{Wysocki2019,LIGO-O3-O3a-catalog}.
By contrast, due to their computational complexity, physics-based models are
customarily explored via Monte Carlo, propagating the results of many randomly selected initial conditions through their
(at times stochastic) evolutionary model.  
Because both these simulations' outputs and GW observations are represented by Monte Carlo samples, a continuous
approximation of (at least) one set of samples is required to produce the net likelihood of the model, given GW
observations.  
In this work, we will approximate the GW likelihood by normal distributions, and use the StarTrack
    binary evolution suite
    \cite{Belczynski2002, Belczynski2008, Belczynski2016, Belczynski2020}; see Appendix \ref{ap:StarTrack} for pertinent inputs.

Following our longer companion study  \cite{st_inference_interp}, we evaluate the (marginal) inhomogeneous Poisson
likelihood for existing GW observations, given each specific population synthesis model. 
We use  the entire simulated merger population ($\sim 10^8$ samples); only a fast approximation like a multivariate
normal makes calculations at this scale possible. 

In this work, we report on a previously-reported and extensively-studied one-parameter family of binary evolution
models, which varies only  the strength of supernova
    natal recoil kicks on newly formed compact remnants. 
All input models are publicly available via
    \url{www.syntheticuniverse.org} \cite{Belczynski2020}.

   Figure  \ref{fig:Application} shows our results for the M13--M19 (sub-model B) family of models.
The solid lines show each model's net likelihood.
The dotted line and shaded region show the results of carefully
    interpolating between these models' results with a Gaussian process,
    accounting for systematics in each evaluated
    likelihood.
Consistent with several previous studies, we find the net likelihood peaks with kick velocities around
$100\unit{km/s}$.  
Nominally, this peak reflects two factors which arise from the model's predictions as weighted by GW search selection effects: the expected number of detections should be consistent with the observed
count; and the distribution of observed masses should be consistent with the models' selection-weighted
predictions.  However, in our simple application, the mass distribution of the M13--M19 models varies relatively little,
so the strong peak in the overall likelihood principally reflects the point where the observed count is close to the
expected count.

\section{Conclusions}
\label{sec:conclude}
Motivated by well-studied analytic approximations, we have introduced a straightforward but powerful technique to approximate existing posterior
distributions via a Gaussian likelihood.  We demonstrated this  approach performs surprisingly  well with just 3 key
dimensions ($\mc,\eta,\chi_{\textrm{eff}}$) for all events reported in GWTC-1 and GWTC-2.  

In subsequent work, we comment on how well this approximation 
    continues to perform when all intrinsic degrees of freedom,
    distance, and tides are included in previously reported-events.   

The increasing degree of precision with which can be characterized may allow for better constrained parameterization
    for additional degrees of freedom in dimensions such as 
    transverse spin for future events.
We also expect to comment on how well this approach performs for 
    such degrees of freedom as well in future work.

This extremely compact and analytically-tractable representation of GW observations will enable end-users to perform robust analyses with many inputs. 

With modest extension to estimate the eigenvalues and eigenvectors of these Gaussians,
    via ansatz or interpolation, 
    this work also provides a straightforward path toward
    generating extremely realistic synthetic GW catalogs.
Such catalogs have wide application to investigate GW population astrophysics.

\begin{acknowledgements}
The authors thank Patricia Schmidt for helpful feedback. 
ROS, VD, and AY are supported by NSF PHY-2012057; ROS is also supported via NSF PHY-1912632 and AST-1909534. DW is supported by NSF PHY-1912649.
This material is based upon work supported by NSF’s LIGO Laboratory which is a major facility fully funded by the
National Science Foundation.
This research has made use of data, software and/or web tools obtained from the Gravitational Wave Open Science Center
(https://www.gw-openscience.org/ ), a service of LIGO Laboratory, the LIGO Scientific Collaboration and the Virgo
Collaboration. LIGO Laboratory and Advanced LIGO are funded by the United States National Science Foundation (NSF) as
well as the Science and Technology Facilities Council (STFC) of the United Kingdom, the Max-Planck-Society (MPS), and
the State of Niedersachsen/Germany for support of the construction of Advanced LIGO and construction and operation of
the GEO600 detector. Additional support for Advanced LIGO was provided by the Australian Research Council. Virgo is
funded, through the European Gravitational Observatory (EGO), by the French Centre National de Recherche Scientifique
(CNRS), the Italian Istituto Nazionale di Fisica Nucleare (INFN) and the Dutch Nikhef, with contributions by
institutions from Belgium, Germany, Greece, Hungary, Ireland, Japan, Monaco, Poland, Portugal, Spain.
The authors are grateful for computational resources provided by the LIGO Laboratory and supported by 	National Science Foundation Grants PHY-0757058 and PHY-0823459.
\end{acknowledgements}

\footnotesize\bibliography{paperexport,references,Bibliography,LIGO-publications,mm-statistics}

\appendix
\section{Estimating StarTrack Model Likelihoods}
\label{ap:StarTrack}

In this section we briefly summarize the StarTrack simulation suite and 
    the methods which allow us to draw conclusions from StarTrack populations
    \cite{Belczynski2002, Belczynski2008, Belczynski2016, Belczynski2020}.
StarTrack evolves binary star systems, and considers a host of physical processes
    including accretion, tidal interactions,
    stellar wind, metallicity, gravitational radiation, magnetic braking,
    and supernova natal recoil kicks \cite{Belczynski2002}.
Populations are drawn at a distribution of redshifts,   
    and are weighted appropriately to create accurate merger rate
    densities for those systems which become compact binaries
    that merge in cosmological time
    \cite{2012ApJ...759...52D,startrack-paper, tides}.
Cosmological postprocessing for StarTrack populations
    in a realistic spread of metallicities allows for the generation of
    a synthetic universe with a physically motivated
    population of compact binary merger rate densities \cite{tides}.
These densities allow us to construct an estimate of the merger rate density  $\rho_i(\mathbf{\lambda})$
    for a given population (e.g., $i$ indexes M13, M14, etc), 
    a physically scaled density function in merger parameters,
    $\mathbf{\lambda}$ \cite{Wysocki2019}.

In order to properly constrain formation parameters
    $\Lambda_i$, for a population $i$,
    we must calculate the joint likelihood for each population:
\begin{align}
    P(\{d_j\} | \Lambda_i) = 
        P(\mu_i | \Lambda_i) \prod\limits_j P(d_j | \Lambda_i)
\end{align}
Here, $\{d_j\}$ denotes a set of gravitational wave observations,
    individually $d_j$.
The joint likelihood is derived from the marginalized event likelihood
    for each observation,
    $P(d_j | \Lambda_i)$
    and the inhomogeneous Poisson likelihood for the total number
    of observations of each type (the rate likelihood),
    $P(\mu_i | \Lambda_i)$ \cite{Wysocki2019}.
In this study, the joint likelihood is dominated by the event likelihood,
    which changes more rapidly than the rate likelihood
    from one population to another.
For a more diverse set of population models,
    both the rate likelihood and event likelihood 
    will be important for constraining formation parameters.

\subsubsection{Rate Likelihood}

The probability that LIGO detection rates would match 
    the expected rate within a population for
    each type of event
$\alpha \in$ \{BBH, BNS, NSBH\} 
is given by the inhomogeneous Poisson likelihood \cite{Wysocki2019}:
\begin{align}
P(\mu_{\alpha, i}|\Lambda_i) = \
    e^{-{\mu_{\alpha,i}}} \frac{\mu_{i,\alpha}^{N_\alpha}}{N_\alpha!}
\end{align}
Here, $\mu_{\alpha,i}$ is the predicted detection rate for a population,
    and $N_{\alpha}$ is the number of published detections of each type
    for a given observing run.
Therefore, the combined rate likelihood is
\begin{align}\label{eq:rate_likelihood}
 P(\mu_i|\Lambda_i) = 
     \prod\limits_{\alpha}
        \bigg[\frac{(\mu_{\alpha, i})^{N_{\alpha}}}{N_{\alpha}!} \bigg]
     e^{-\sum\limits_{\alpha} \mu_{\alpha,i}} =
        K_ie^{-\mu_i}
 \end{align}

In our own work \cite{st_inference_interp},
    we predict the expected detection rates for each population
    by the method outlined by Dominik et al. \cite{tides}
\begin{align}\label{eq:LIGOSensitivity}
\mu_{det} = T_{\mathrm{obs}}\sum \frac{\mathrm{SFR}}{\Delta M_f}p_{\mathrm{det}} \frac{1}{1 + z} 
    \frac{dV_c}{dz} \frac{dz}{dt} \Delta t
\end{align}
$\mathrm{SFR}/\Delta M_f$ is a weighted star formation rate,
    and in our work is equivalent to $\rho_i(\mathbf{\lambda})$.
$dV_c / dz$ is the differential comoving volume
    according to a conventional Planck2015 cosmology \cite{Ade:2015xua}.
Additionally, $dz/dt = (1 + z) H$, and $\Delta t = 100\mathrm{Myr}$ 
    is the cosmological time step.
The detector sensitivity, $p_{\mathrm{det}}(w)$, is interpolated from results
    tabulated by other groups (\url{https://pages.jh.edu/~eberti2/research/})
    \cite{tides}.
The signal-to-noise ratio is 
    interpolated using a sparse Gaussian process regression
    method, and evaluated for each merger sample \cite{st_inference_interp}.
All cosmological calculations are implemented with \textsc{astropy} \cite{astropy}.

\subsubsection{Event Likelihood and Marginalization}

The marginalized event likelihood $P(d_j | \Lambda_i)$ must be calculated 
    using the associated event likelihood 
    $\mathcal{L}_j(\mathbf{\lambda})$
    and normalized merger rate density function for each population,
    $\bar{\rho}(\mathbf{\lambda})$.
For each of the $~10^8$ sample binary mergers (samples are indexed by $k$)
    in a population with formation parameters $\Lambda_i$,
    each event likelihood must be calculated
    $\mathcal{L}_j (\lambda_{i,k}) = P(d_j | \lambda_{i,k} \Lambda_i)$.
With the same indexing, the normalized merger rate densities for each sample
    are $\omega_{i,k}$.
That evaluation would not be possible without an approximate model for
    the event likelihood which could be evaluated in 
    a small fraction of a second.
We use Normal approximations to the likelihood, such as those which
    appear in this publication.

The marginalization is as follows 
\begin{align}\label{eq:likelihood_synthesis}
P(d_j| \Lambda_i) = \int\limits_{\{\mathbf{\lambda}\}}
        P(d_j|\mathbf{\lambda},\Lambda_i)P(\mathbf{\lambda}|\Lambda_i) 
        \mathrm{d}\mathbf{\lambda} =
        \int\limits_{\{\mathbf{\lambda}\}}
    \bar{\rho}_i(\mathbf{\lambda}) \mathcal{L}_j(\mathbf{\lambda})
        \mathrm{d}\mathbf{\lambda}
\end{align}
This brings us to our discrete expression \cite{Wysocki2019} for the joint likelihood:
\begin{align}\label{eq:wysocki-likelihood}
P(\{d_j\} | \Lambda_i) \propto
    K_{\mathrm{rate},i} \mathrm{e}^{-\mu_{\mathrm{det}, i}} 
    \prod\limits_j \bigg[ \sum\limits_k \mathcal{L}_j(\lambda_{i,k}) \omega_{i,k} \bigg]
\end{align}

\end{document}